\documentclass[%
 reprint,
 amsmath,amssymb,
 aps,
]{revtex4-2}

\usepackage{graphicx}% Include figure files
\usepackage{dcolumn}% Align table columns on decimal point
\usepackage{bm}% bold math
\usepackage[T1]{fontenc} % if needed
\usepackage{cancel}
\usepackage{hyperref}
\begin{document}

\preprint{APS/123-QED}

\title{Reparametrization Symmetry of Local Entropy Production on a Dynamical Horizon}% Force line breaks with \\

\author{Sayantani Bhattacharyya}
 \email{sayanta@niser.ac.in}
 \affiliation{%
 School of Physical Sciences, National Institute of Science Education and Research\\
 An OCC of Homi Bhabha National Institute, Jatni-752050, India
}
\author{Pooja Jethwani} %\footnote{We are deeply saddened to lose our friend and collaborator, Pooja Jethwani. She was an active collaborator in this project and had seen its completion before the end of our time together.}}%
 \thanks{We are deeply saddened to lose our friend and collaborator, Pooja Jethwani. She was an active collaborator in this project and had seen its completion before the end of our time together.}
 \affiliation{%
 School of Physical Sciences, National Institute of Science Education and Research\\
 An OCC of Homi Bhabha National Institute, Jatni-752050, India
}
\author{Milan Patra}
 \email{milan.patra@niser.ac.in}
 \affiliation{%
 School of Physical Sciences, National Institute of Science Education and Research\\
 An OCC of Homi Bhabha National Institute, Jatni-752050, India
}
\author{Shuvayu Roy}
 \email{shuvayu.roy@niser.ac.in}
\affiliation{%
 School of Physical Sciences, National Institute of Science Education and Research\\
 An OCC of Homi Bhabha National Institute, Jatni-752050, India
}%


%\affiliation{
%\affiliation{
%\affiliation{%


\date{\today}% It is always \today, today,

\begin{abstract}
Recently, it has been shown that for a dynamical black hole in any higher derivative theory of gravity, one could construct a spatial entropy current, characterizing the in/outflow of entropy at every point on the horizon, as long as the dynamics of the amplitude is small enough. However, the construction is very much dependent on how we choose the spatial slicing of the horizon along its null generators. In this note, we have shown that though both the entropy density and the spatial entropy current change non-trivially under a reparametrization of the null generator,  the net entropy production, which is given by the `time' derivative of entropy density plus the divergence of the spatial current is invariant. We have explicitly verified this claim for the particular case of dynamical black holes Einstein-Gauss-Bonnet theory. 
%\begin{description}
%\end{description}
\end{abstract}

\maketitle

%\tableofcontents

\section{Introduction}\label{sec:intro}
It is expected that a low energy effective description of any UV complete theory of gravity will typically have higher derivative corrections to Einstein's theory of gravity. We could further expect that such a corrected theory of gravity would admit a classical limit and also black-hole-type classical solutions with curvature singularity shielded by an event horizon.

It is well-known that black hole solutions in two derivative theories of gravity are analogous to  large thermodynamic objects with many underlying degrees of freedom \cite{Bardeen:1973gs, Bekenstein:1973ur, Hawking:1973uf}. One could associate temperature, energy (and other conserved charges) and entropy to every black hole geometry which satisfies all the laws of thermodynamics.

But, for black holes in higher derivative theories, we still do not have a complete understanding of all their thermodynamic properties. Specifically, in a dynamical situation, we do not know which geometric property of the black hole to identify with the system's entropy so that it satisfies the second law of thermodynamics. 

However, we know how to construct entropy geometrically even in any  higher derivative theory if the black hole is stationary \cite{WaldEnt, Iyer:1994ys}. This entropy, by construction, satisfies the first law of thermodynamics. But if the black hole is dynamic, several corrections that necessarily vanish in a stationary situation and do not affect the first law could be added to this entropy expression \cite{Jacobson:1993xs, Jacobson:1993vj, Jacobson:1995uq}.

Recently in \cite{Wall}, the author has fixed some of these ambiguities in the expression of entropy by studying gravitational dynamics of very small amplitude so that any term nonlinear in this amplitude could be neglected. Then in \cite{our1912} and \cite{our2105}, the authors have constructed a spatial current whose divergence could be identified with  the entropy in/out flow in any infinitesimal subregion of the horizon. Using the entropy density and the spatial entropy current, one could restate the second law in an ultra-local fashion where entropy is produced in every infinitesimal subregion of the horizon for a generic dynamics slow enough so that all the higher derivative corrections could be treated perturbatively.

However, this construction of entropy density and the current relies on a very specific choice of the coordinate system where the affine parameter along the null generator of the horizon is one of the coordinates. Now it is possible to reparameterize the null generators of the horizon in a nontrivial way without affecting the affineness of the parameters. The expressions for both the entropy density and the spatial current change under this reparametrization. But we  expect the net entropy production, given by the `time' derivative of the entropy density plus the divergence of the spatial current, should be something physical and therefore, independent of our choice of affine parameters.
 
In this note, our goal is to verify the above expectation for the special case of Gauss-Bonnet theory where both the entropy density and the current have been explicitly computed in  \cite{our1912}.
 
We have found that under this transformation, the ‘time’ derivative of the entropy density as well as the divergence of the spatial entropy current change individually in a very nontrivial way; however they precisely cancel each other. Apart from being a consistency check for the results described in \cite{our1912}, it also says why a spatial entropy current is necessary to make the laws of entropy production independent of our choices of coordinates.
 
Though at the moment all the calculations are linear in the amplitude of the dynamics, we eventually would like to have some construction of entropy density and the entropy current that satisfy the first and the second law at all nonlinear orders and if possible, without using any perturbation. Now, in any such construction, a full knowledge of the underlying symmetries might turn out to be very useful. The requirement that the entropy current and the density must transform in such a way so that the net entropy production has some particular symmetry could be constraining for all the nonlinear terms. \footnote{In \cite{Hollands:2022fkn} which came up shortly after our work, the authors have included an elaborate discussion on this issue.} In other words, it would be very interesting if, instead of verifying the symmetry in a particular theory, we could use it to predict some relation between the structure of entropy density and the spatial entropy current in a theory independent manner. We expect that our explicit computation in the simple case of Gauss-Bonnet theory would help us to gain experience for further progress in this direction.

The note is organized as follows.
In section \ref{sec:setup}, we have described the setup of our problem, including the choice of our coordinates adapted to the horizon.  In section \ref{sec:symmetry} we have described the reparametrization symmetry. In section \ref{sec:verify} we have explicitly verified that the entropy density and the entropy current do maintain this symmetry in the particular case of Gauss-Bonnet theory. Finally in section \ref{sec:conclude}, we have concluded. The details of the calculation are explained in several appendices.

\section{Set up}\label{sec:setup}
In this section, we shall briefly review the coordinate system used in the analysis of \cite{our1912} and the expression for entropy current and entropy density for the Gauss-Bonnet theory. 

\subsection{Coordinate system}\label{subsec:coordinate}
As mentioned before, the geometry we are considering is of the black-hole type containing a codimension one null surface as the horizon. The coordinate system is constructed with the horizon being the base i.e. we first choose  $(D-1)$ coordinates on the horizon. Let $\partial_v$ is the generator of the horizon which is a null geodesic with $v$ being the affine parameter and $x^a,~~~\{a=1,\cdots,D-2\}$ are the spatial coordinates along the constant $v$ slices of the horizon. So $\{v,x^a\}$ together constitute a coordinate system on the horizon.

Once the coordinates on the horizon are fixed, we shoot off affinely parametrized null rays $\partial_r$, making specific angles with horizon coordinates. The affine parameter $r$ along these rays is a measure of the distance away from the horizon. The angles are chosen so that the inner product between $\partial_r$ and $\partial_v$ on the horizon is $1$ and the same between $\partial_r$ and $\partial_a$s are zero. After imposing all these conditions, the metric takes the following form (see \cite{our1912} for more details)
\begin{equation}\label{eq:metric}
\begin{split}
ds^2 &= 2 ~dv~ dr -r^2 ~X(r,v,x^a)~dv^2 \\
&+ 2~ r~\omega_a(r,v,x^b)~ dv~dx^a + h_{ab}(r,v,x^a)~dx^a~dx^b
\end{split}
\end{equation}

\subsection{Gauss-Bonnet theory}\label{subsec:gaussbonnet}
We are considering a theory of pure gravity with  maximum four derivatives.  We shall be even more specific in choosing the theory; we'll work with the Gauss-Bonnet theory of gravity with the following Action.
 \begin{equation}\label{eq:GaussBonnetAction}
 S =\int d^Dx\sqrt{-G}\left[R+\alpha^2\left(R^2-4R^{\mu\nu}R_{\mu\nu}+R^{\mu\nu\rho\sigma}R_{\mu\nu\rho\sigma}\right)\right]
 \end{equation}
 Here $R$, $R_{\mu\nu}$ and $R_{\mu\nu\rho\sigma}$ are the Ricci scalar, Ricci tensor, and Riemann tensor\footnote{ According to our convention, 
 \begin{equation*}\label{eq:riemann}
 \begin{split}
 &R\equiv g^{\mu\nu}R_{\mu\nu},~~~~R_{\mu\nu} \equiv {R^\rho}_{\mu\rho\nu}\\
 &{R^\mu}_{\nu\rho\sigma}\equiv  \partial_\rho\Gamma^\mu_{\nu\sigma} - \partial_\sigma\Gamma^\mu_{\nu\rho} + \Gamma^\mu_{\rho\alpha}\Gamma^\alpha_{\nu\sigma} -\Gamma^\mu_{\sigma\alpha}\Gamma^\alpha_{\rho\nu} 
 \end{split}
 \end{equation*}} of the full spacetime respectively. All raising and lowering of indices have been done using the bulk metric $g_{\mu\nu}$.
 
The entropy density ($J^v$) and the entropy current ($J^a$) on the horizon have the following structure
\begin{equation}\label{eq:entcur}
 \begin{split}
 J^v &=\left(1+2 \alpha^2\mathcal R\right)\\
    J^a &=\alpha^2\left[-4\nabla_b K^{ab}+4\nabla^a K\right]
 \end{split}
\end{equation}
Here ${\cal R}$ is the intrinsic Ricci scalar of the constant $v$ slices of the horizon (i.e., the Ricci scalar computed using the metric $h_{ab}$). $K_{ab}$ is the extrinsic curvature of the null horizon, and $\nabla_a$ is the covariant derivative with respect to $h_{ab}$
\begin{equation}\label{eq:not}
 \begin{split}
 K_{ab} \equiv{\frac{1}{2}}\partial_v h_{ab},~~~K \equiv h^{ab} K_{ab}
 \end{split}
 \end{equation}
The sole reason for choosing this theory is its simplicity. Despite being a four derivative theory, the equation of motion remains two derivative and both the entropy density and the current could be constructed entirely from $h_{ab}$ and its $v$ and $x^a$ derivatives evaluated on the horizon, which simplifies our task to a large extent. However, we must emphasize that the symmetry that we are going to describe in the next section is expected to hold in any higher derivative theory of gravity.
 
\section{Symmetry}\label{sec:symmetry}
In section \ref{sec:setup}, we have chosen a coordinate system adapted to the horizon so that the metric takes the form as described in equation \eqref{eq:metric}. However, this form does not fix the coordinates completely, some residual gauge freedom is still left and both the entropy density and entropy current do change non-trivially under this unfixed coordinate freedom. 

On the other hand, as we have explained in the introduction, the expression $$\left[\frac{1}{\sqrt{h}} \partial_v (\sqrt{h} J^v)  + \nabla_i J^i\right]$$ (where $J^v$ and $J^i$ are the entropy density and the spatial entropy current, respectively) is related to the local entropy production along every point of the dynamical horizon and therefore, we expect it to be invariant under the reparametrization of the null generators. 

In this section, we shall first describe this residual freedom of coordinate transformation that is not fixed by our choice of gauge. Next, we shall use the details of this transformation to make our intuition about `invariance' more precise. 

\subsection{Reparametrization of the null generator}

The starting points in setting up our bulk coordinate system are the affinely parametrised null generators of the horizon and the  coordinates along its spatial  slices.  Once we fix the horizon coordinates, our gauge conditions uniquely fix the coordinates along the bulk. It follows that the residual symmetry that we are going to discuss here must involve  a transformation of the horizon coordinates maintaining the affineness of the null generators.
For convenience,  let us use a bar on all the coordinates of the horizon to distinguish them from the bulk coordinates. For example, $\{\bar v,\bar x^a\}$ will denote the affine parameter along the null generator and spatial coordinates along the constant $\bar v$ slices of the horizon only. 

Now an affine parameter will remain an affine parameter if we scale it in a $\bar v$ independent manner. So we shall consider the following transformation on the horizon ($r=0$ hypersurface).
\begin{equation}\label{repeq1}
\bar v\rightarrow\bar \tau = \bar v ~e^{-\zeta(\bar x^a)},~~~\bar x^a\rightarrow \bar y^a = \bar x^a
\end{equation}
As mentioned before, both $\bar v$ and $\bar\tau$ are affine parameters along the null generators of the horizon. However constant $\bar v$ slices are not the same as the constant $\bar \tau$ slices. In other words, the tangent vectors along the constant $\bar v$ slices given as $\bar\partial_a^{(x)}$ are different from the tangent vectors $\bar\partial^{(y)}_a$ along the constant $\bar \tau$ slices. They are related as follows

\begin{equation}\label{repeq2}
\bar\partial_a^{(x)} = \bar \partial_a^{(y)} - \left(\frac{\partial\zeta} {\partial\bar y^a}\right)\bar\tau\partial_{\bar\tau}
\end{equation}

Since the tangent vectors on the horizon change under this transformation, we need to transform the $ r$ coordinate also so that the tangents along the constant $\{\tau,y^a\}$ lines (or the coordinate vectors pointing away from the horizon) maintain the same angle with the coordinate vectors along the horizon. This will firstly lead to a redefinition of the $r$ coordinate and also, it will correct the coordinate transformation \eqref{repeq2} as one goes away from the horizon.

\begin{equation}\label{eq:fullcoord}
\begin{split}
v &= e^{\zeta(y)}\tau \left[1 + \sum_{n=1}(\rho~ \tau)^n ~V_{(n)} (\tau, \vec y)\right]\\
r &= e^{-\zeta(y)}\rho \left[1 + \sum_{n=1}(\rho~ \tau)^n ~R_{(n)} (\tau, \vec y)\right]\\
x^a &= y^a +\sum_{n=1}(\rho~ \tau)^n ~Z^a_{(n)} (\tau, \vec y)
\end{split}
\end{equation}

Let us briefly motivate the  choice of above ansatz .\\
 As mentioned before, the coordinate transformation is generated due to the scaling function $\zeta(\bar y)$ defined only on the horizon and once this horizon function is given, the rest of the coordinates throughout the bulk are uniquely determined by our gauge condition. Clearly it is impossible to solve these gauge conditions exactly for a generic space time. But the problem is very well-suited for a near horizon expansion since geometrically our choice of gauge is a two-step process where we first choose coordinates on the horizon and then shoot out null geodesics with precise angles to extend them away from the horizon. 
 
 As it is often true with perturbative expansions, our ansatz  also involves few conventions and assumptions.
First note that each of the expansion coefficients ($V_{(n)}, R_{(n)}$ and $Z^a_{(n)}$), including the function $e^{\pm\zeta}$ strictly speaking should depend only on the horizon coordinates $\{\bar\tau,\bar y^a\}$. Whenever we are writing them as functions of bulk coordinates $\{\tau,y^a\}$ it involves an extension of these functions to the bulk, which is rather arbitrary. It is always possible to redefine  the expansion coefficients at any given order by adding functions that vanish on the horizon without affecting the lower order coefficients. Similarly  $\zeta$ itself might admit a power series expansion in distance from the horizon (in fact if we choose to write $\zeta(y^a)$ in terms of $\{x^a\}$ coordinates this will happen). However, such redefinition, geometrically does not mean that we are choosing new curves for coordinate axes, since we know all coordinates are uniquely determined by our gauge choice once we fix the coordinates on the horizon. This is simply a rearrangement redundancy that is built into our perturbative technique of solving the gauge choices.  However, here we have chosen the most naive bulk extension of all these horizon quantities by simply replacing all the $\{\bar\tau,\bar y^a\}$ dependence with bulk coordinates $\{\tau,y^a\}$ (which need not be the simplest choice in terms of the final form of the expansion coefficients).

 Next we shall come to the second unusual choice we made in our ansatz. A near horizon expansion in our coordinates simply means an expansion in powers of $\rho$ (and not in the powers of the product $(\rho\tau)$ as we have done here). Though note that  there is no loss of generality in expanding in the powers of the product $(\rho\tau)$ if we keep the $\tau$ dependence in the expansion coefficients completely free. The reason behind this choice of expansion parameter is related to equilibrium (stationary) horizons. We know that in stationary black holes the radial dependence of the metric components is always through the boost-invariant product $(\rho\tau)$ or $(rv)$ \cite{our1912}. This will be true provided the coordinate transformation has the structure as described above with coefficient functions independent of $\tau$ coordinates. In other words in our $(\rho\tau)$ expansion, the expansion coefficients will depend on $\tau$ only when the horizon is evolving with time, thus enabling us to clearly distill out the effect of dynamics from that of stationary case.\\
 
 Fortunately all these subtle issues about the form of the coordinate transformation turn out to be completely irrelevant for the present analysis of Einstein Gauss Bonnet gravity. For this theory both the entropy density and entropy current  are entirely constructed out of the induced spatial metric of the horizon (denoted as $h_{ab}$) and its derivative along the tangents of the horizon (i.e., $\partial_a$ and $ \partial_v$ only and no $\partial_r$). Here we do not need to know the metric components away from the horizon and therefore there is no need to determine  the coordinate transformation for nonzero $\rho$.\footnote{Higher order corrections to the metric coefficients are going to be computed in an upcoming work.} The induced metric on the horizon remains invariant under the reparametrization as
\begin{equation}\label{eq:chmetric2}
    \begin{split}
        \Tilde{h}_{ij} &= h_{ij}  + \mathcal{O}(r)
    \end{split}
\end{equation}

\subsection{Why we expect this transformation to be a symmetry}

Here, we shall present a heuristic argument of why we expect such a symmetry to be there in the first place. The argument is very similar to what one uses to prove `the physical process version of the first law.'

Following the setup in \cite{Jacobson:1995uq}, consider a stationary black hole. The horizon is a Killing horizon in the absence of any perturbation. At some Killing time $t_0$, matter fields are perturbed. If we treat the amplitude of the field perturbation as of (${\cal O}(\delta)$), then typically, the fluctuation in the matter stress tensor would be of order ${\cal O}(\delta^2)$ and the same would be the order of the metric fluctuation (which, at later sections, has been denoted as $\epsilon\sim\delta^2$). It follows that the local entropy production $S_p\equiv \bigg[{1\over\sqrt h}\partial_v\left({\sqrt h} ~J^v\right) + \nabla_iJ^i\bigg]$, which is constructed solely out of metric fluctuation, is also of order ${\cal O}\left(\delta^2\right)$. Note that the Killing equation will remain true up to order ${\cal O} (\delta)$ and therefore to compute the leading order (${\cal O}(\delta^2)$) expression for the entropy production, it makes sense to integrate $S_p$ between two constant `Killing time' slices of the horizon, namely initial equilibrium (at `Killing time $t=-\infty$) to final equilibrium (at Killing time $t=\infty$). Now we could relate the `Killing time' to the affine parameter of the null generators where $t=-\infty$ will correspond to $v=0$, and $t=\infty$ will correspond to $v=\infty$ (see \cite{Jacobson:1995uq} for the details). So, the net entropy production could be expressed as \cite{Jacobson:1995uq, our1912, our2105, Gao:2001ut, Amsel:2007mh, Bhattacharjee:2014eea, Chakraborty:2017kob, Chatterjee:2011wj} 
\footnote{We thank the referee for clarifying this point to us.}

\begin{equation}\label{eq:diffent}
\begin{split}
\Delta S &= \int_{0}^{\infty} dv\int_{\Sigma_v}d^n\vec x~\sqrt{h}\left[\frac{1}{\sqrt{h}}\partial_v\left(\sqrt{h} ~J^v\right) + \nabla_i J^i\right] \\
&= S_{Equilibrium_2}-S_{Equilibrium_1}
\end{split}
\end{equation}
where $\Sigma_v$ is the constant $v$ slices of the horizon and $n=D-2$.

But the total entropy in an equilibrium or for a stationary black hole is unambiguously defined through Wald entropy, which is independent of how we parametrize the null generators of the horizon and the same must be true of their difference. Now under the reparametrization that we are discussing, the measure of the above integration changes as
$$\sqrt{h}~dv~d^n\vec x = e^{\zeta(y)} \sqrt{h}~d\tau~d^n\vec y$$
If we want $\Delta S$ to be invariant under the reparametrization of the null generators, then the expression $\left[\frac{1}{\sqrt{h}}\partial_v\left(\sqrt{h} ~J^v\right) + \nabla_i J^i\right]$ , once written in terms of quantities defined in $\{\tau, \vec y\}$ coordinates, must have an overall factor of $e^{-\zeta}$.
\begin{equation}\label{eq:transform}
\left[\frac{1}{\sqrt{h}}\partial_v\left(\sqrt{h} ~J^v\right) + \nabla_a J^a\right]=e^{-\zeta}\Big[\frac{1}{\sqrt{\tilde h}} \partial_\tau(\sqrt{\tilde h} \tilde J^\tau)+\tilde\nabla_a \tilde J^a\Big]
\end{equation}
Here the LHS is expressed in $\{v,\vec x\}$ coordinates and RHS is in $\{\tau, \vec y\}$ coordinates.\\

Now we shall come to an algebraic reason why the expression for net entropy production should transform exactly as predicted in equation \eqref{eq:transform}. We shall restrict this discussion to the theories of pure gravity.

The key equation that leads to the entropy current on the horizon is the following
\begin{equation}
   E_{vv}|_{r=0}=\partial_v\Big[\frac{1}{\sqrt{h}} \partial_v(\sqrt{h} J^v)+\nabla_a J^a\Big],\label{eq:key}
   \end{equation}
Here $E_{vv}$ is the $(vv)$ component of the equation of motion. This is a component of a covariant tensor and therefore, we know how it transforms under the above coordinate transformation for every possible gravity action. On the horizon (i.e., at $\rho=0$ hypersurface) the transformation becomes particularly simple.
   \begin{equation}\label{eq:evvtrans}
   \begin{split}
   E_{vv}\vert_{r=0}= e^{-2\zeta}E_{\tau\tau}\vert_{r=0}
   \end{split}
   \end{equation}
   
Now in $\{\rho,\tau, y^a\}$ coordinates the metric has the same form as in equation \eqref{eq:metric}. Therefore $E_{\tau\tau}$  could also be expressed as in equation \eqref{eq:key} for some $\tilde J^\tau$ and $\tilde J^a$. 
 $$E_{\tau\tau}\vert_{r=0} = \partial_\tau\Big[\frac{1}{\sqrt{\tilde h}} \partial_\tau(\sqrt{\tilde h} \tilde J^\tau)+\tilde\nabla_a \tilde J^a\Big]$$
 Note $\tilde J^\tau$ and $\tilde J^a$ are not components of covariant tensors on bulk space and therefore they do not transform in any well-defined way. But combining the above equation with equations \eqref{eq:evvtrans} and \eqref{eq:key} we get the following prediction.
 \begin{equation}\label{eq:predic}
 \begin{split}
 E_{vv}\vert_{r=0} &= \partial_v\Big[\frac{1}{\sqrt{h}} \partial_v(\sqrt{h} J^v)+\nabla_a J^a\Big]\\
 &= e^{-\zeta}~\partial_\tau\Big[\frac{1}{\sqrt{h}} \partial_v(\sqrt{h} J^v)+\nabla_a J^a\Big]\\
 &= e^{-2\zeta}E_{\tau\tau} \\
 &=e^{-2\zeta}~\partial_\tau\Big[\frac{1}{\sqrt{\tilde h}} \partial_\tau(\sqrt{\tilde h} \tilde J^\tau)+\tilde\nabla_a \tilde J^a\Big]\\
 \Rightarrow \frac{1}{\sqrt{h}} \partial_v(\sqrt{h} & J^v)+\nabla_a J^a=e^{-\zeta}\Big[\frac{1}{\sqrt{\tilde h}} \partial_\tau(\sqrt{\tilde h} \tilde J^\tau)+\tilde\nabla_a \tilde J^a\Big]
 \end{split}
 \end{equation}
In the last line, both LHS and RHS (up to the factor of $e^{-\zeta}$)  are related to the net entropy production in the two coordinate systems discussed here.  It follows that though the entropy density and the entropy current might change in a very nontrivial way with several terms dependent on derivatives of $\zeta$, in the final expression of entropy production, they must cancel, leaving just an overall $e^{-\zeta}$ factor.
Further, the equation \eqref{eq:predic} also says that this nontrivial cancellation must be true in all higher derivative theories of gravity. In the next section, we shall verify this claim in the simplest case of Gauss-Bonnet theory.
\footnote{It might seem that the heuristic justification provided at the very beginning of this subsection is not very different from the algebraic one involving $E_{vv}$. Indeed, if we follow the argument presented in \cite{Jacobson:1995uq}, we see that at linearized order, the net entropy production has been first related to the integration of the $\{vv\}$ component of the matter stress tensor and then by the equation of motion is related to the integration of $E_{vv}$. So, the covariance of the integrand in (eqn 9) is effectively the same as the covariance of $E_{vv}$  at least in this order. However, the covariance of the integrand has a scope for further generalization if we want to extend this construction to higher orders in amplitude expansion. Following \cite{Hollands:2022fkn}, we could see that as we go in higher order, this local entropy current can no longer be derived just from $E_{vv}$, but the other components of $E_{\mu\nu}$ also contribute, and it becomes quite complicated to figure out the net transformation property of this combination of equations. However, if we expect the ultra-local form of entropy production to be valid at higher orders, then there must be an integration formula for $\Delta S$, and the integrand must transform in a covariant manner once the corrections to Killing equations have been appropriately taken care of.}

\section{Verification for Gauss-Bonnet Theory}\label{sec:verify}
In this section, for the special case of Gauss-Bonnet theory, we would like to explicitly verify whether the local entropy production on the horizon transforms the way we have predicted in the previous sections. We know 
\begin{equation}
   E_{vv}|_{r=0}=\partial_v\Big[\frac{1}{\sqrt{h}} \partial_v(\sqrt{h} J^v)+\nabla_a J^a\Big],\label{eq2.1}
\end{equation}
where 
\begin{align}
    J^v &=1+2\alpha^2  \mathcal{R},\label{eq2.2}\\
    J^a &=\alpha^2\left[-4\nabla_b K^{ab}+4\nabla^a K\right]\label{eq2.3}
\end{align}
On the horizon, the reparametrization we are considering is the following
\begin{align}
    v&=\tau\ e^{\zeta(y)},\label{eq2.4}\\
    x^a&=y^a.\label{eq2.5}
\end{align}
Clearly the ${\cal O}(\alpha^0)$ piece (contribution from Einstein gravity) in $J^v$ does not transform. So  now we have to determine how the order ${\cal O}(\alpha^2)$ pieces of $J^v$ and $J^a$ transform. Both of them will receive non-trivial shifts generated by derivatives of the function $\zeta(\vec y)$. But these
shifts will be such that eventually in the expression of $\Big[\frac{1}{\sqrt{h}} \partial_v(\sqrt{h} J^v)+\nabla_a J^a\Big]$ they will precisely cancel up to a factor of overall $e^{-\zeta}$.\\
Now we shall first  describe how all the relevant quantities individually transform under this reparametrization.

The derivatives transform as
\begin{align}
    \partial_v&=e^{-\zeta(y)}\partial_{\tau},\label{eq2.6}\\
    \partial_a &=\Tilde{\partial}_a -(\Tilde{\partial}_a \zeta)\tau\partial_\tau.\label{eq2.7}
\end{align}
The Christoffel connection transforms as 
\begin{equation}
    \begin{split}
         \Gamma _{a,bc}&=\frac{1}{2}(\partial_b h_{ac}+\partial_c h_{ab}-\partial_a h_{bc}),\\
    &=\Tilde{\Gamma} _{a,bc}-\tau (\xi_b \Tilde{K}_{ac}+\xi_c \Tilde{K}_{ab}-\xi_a \Tilde{K}_{bc}),\label{eq2.8}
    \end{split}
\end{equation}
 where, 
\begin{align}
    \xi_a&=\partial_a\zeta=\Tilde{\partial}_a\zeta,\label{eq2.9}\\
     \Tilde{K}_{ab}&=\frac{1}{2}\partial_\tau h_{ab}.\label{eq2.10}
\end{align}
The Ricci scalar is given as 
\begin{equation}
  \Tilde{\mathcal{R}}=(h^{ad} h^{bc}-h^{ac} h^{bd})(\Tilde{\partial}_d\Tilde{\Gamma}_{a,bc}-h^{pq}\Tilde{\Gamma}_{p,ad}\Tilde{\Gamma}_{q,bc})\label{eq2.11}.
     \end{equation}
    Under the change of coordinates, the Ricci Scalar transforms as
    \begin{equation}
    \begin{split}
      \mathcal{R} &=\Tilde{\mathcal{R}} +2(h^{ad} h^{bc}- h^{ac} h^{bd})\\
      [&(-\tau)\{\xi_{bd}+(\xi_b\Tilde{\partial}_d+\xi_d\Tilde{\partial}_b)-\xi_b\xi_d\} \Tilde{K}_{ac}\\
        &+\tau \Tilde{\Gamma}^p_{ad} (\xi_b \Tilde{K}_{pc}+\xi_c \Tilde{K}_{pb}- \xi_p \Tilde{K}_{bc})\\
        &+\tau^2 \xi_b \xi_d \partial_{\tau} \Tilde{K}_{ac}]\label{eq2.12}.
   \end{split}
     \end{equation}
    This implies that the order ${\cal O}(\alpha^2)$ piece of the entropy density transforms as
     \begin{equation}
     \begin{split}
            J^v&=2\mathcal{R} =2\Tilde{\mathcal{R}}\\
            &+4(h^{ad} h^{bc}- h^{ac} h^{bd})[(-\tau)\{\xi_{bd}+(\xi_b\Tilde{\partial}_d+\xi_d\Tilde{\partial}_b)-\xi_b\xi_d\} \Tilde{K}_{ac}\\
        &+\tau \Tilde{\Gamma}^p_{ad} (\xi_b \Tilde{K}_{pc}+\xi_c \Tilde{K}_{pb}- \xi_p \Tilde{K}_{bc})
        +\tau^2 \xi_b \xi_d \partial_{\tau} \Tilde{K}_{ac}].\label{eq2.13}
     \end{split}
     \end{equation}
   Now we know that  $J^\tau\vert_{{\cal O}(\alpha^2)}\equiv 2\Tilde{R} $, then
     \begin{equation}
     \begin{split}
        \frac{1}{\sqrt{h}} &\partial_v(\sqrt{h} J^v)\vert_{{\cal O}(\alpha^2)}\\
        &=  e^{-\zeta} \frac{1}{\sqrt{h}} \partial_\tau\mathcal(\sqrt{h}J^\tau)\vert_{{\cal O}(\alpha^2)}\\
        &+ 4e^{-\zeta}(h^{ad} h^{bc}-h^{ac} h^{bd})\\
        [&-(\xi_{bd} \Tilde{K}_{ac})-(\xi_b\Tilde{\partial}_d+\xi_d\Tilde{\partial}_b) \Tilde{K}_{ac}\\
         &+\Tilde{\Gamma}^{p}_{ad}(\xi_b \Tilde{K}_{pc}+\xi_c \Tilde{K}_{pb}-\xi_p \Tilde{K}_{bc})\\
         &-\tau\{\xi_{bd}+(\xi_b\Tilde{\partial}_d+\xi_d\Tilde{\partial}_b)\}(\partial_\tau \Tilde{K}_{ac})\\
         &+\tau \Tilde{\Gamma}^p_{ad}(\xi_b \partial_\tau \Tilde{K}_{pc}+\xi_c \partial_\tau \Tilde{K}_{pb}-\xi_p\partial_\tau \Tilde{K}_{bc})\\
         &+\xi_b\xi_d \Tilde{K}_{ac}
         +3\tau \xi_b\xi_d\partial_\tau \Tilde{K}_{ac}+\xi_b\xi_d \tau^2\partial_\tau^2 \Tilde{K}_{ac}]\\
         &+\mathcal{O}(\epsilon^2).\label{eq2.14}
     \end{split}
     \end{equation}

         The entropy current is given as
     \begin{equation}
         J^a=-4(h^{ad}h^{bc}-h^{cd} h^{ab})\nabla_b K_{cd},\label{eq2.15}
     \end{equation}
     hence
     \begin{equation}
         \nabla_a J^a=-4(h^{ad}h^{bc}-h^{ac} h^{bd})\nabla_b \nabla_d K_{ac}.\label{eq2.16}
     \end{equation}
     The extrinsic curvature in the two coordinate systems are related as 
     \begin{equation}
         K_{ac} =e^{-\zeta}\Tilde{K}_{ac} \label{eq2.17}.
     \end{equation}
     This implies
     \begin{equation}
         \nabla _dK_{ac} =e^{-\zeta}[\Tilde{\nabla}_d \Tilde{K}_{ac}-\xi_d(\Tilde{K}_{ac}+\tau\partial_\tau \Tilde{K}_{ac})]\\\label{eq2.18}
     \end{equation}
     \begin{equation}
         \begin{split}
              \nabla_b \nabla _d K_{ac} =e^{-\zeta}[&\Tilde{\nabla}_b\Tilde{\nabla}_d \Tilde{K}_{ac}-(\xi_{bd} \Tilde{K}_{ac})-(\xi_b\Tilde{\partial}_d+\xi_d\Tilde{\partial}_b) \Tilde{K}_{ac}\\
         &+\Tilde{\Gamma}^{p}_{ad}(\xi_b \Tilde{K}_{pc}+\xi_c \Tilde{K}_{pb}-\xi_p \Tilde{K}_{bc})\\
         &-\tau\{\xi_{bd}+(\xi_b\Tilde{\partial}_d+\xi_d\Tilde{\partial}_b)\}(\partial_\tau \Tilde{K}_{ac})\\
         &+\tau \Tilde{\Gamma}^p_{ad}(\xi_b \partial_\tau \Tilde{K}_{pc}+\xi_c \partial_\tau \Tilde{K}_{pb}-\xi_p\partial_\tau \Tilde{K}_{bc})\\
         &+\xi_b\xi_d \Tilde{K}_{ac}
         +3\tau \xi_b\xi_d\partial_\tau \Tilde{K}_{ac}+\xi_b\xi_d \tau^2\partial_\tau^2 \Tilde{K}_{ac}]\\
         &+\mathcal{O}(\epsilon^2).\label{eq2.19}
         \end{split}
     \end{equation}

 Hence, the divergence of Entropy current transforms as 
 \begin{equation}
 \begin{split}
     \nabla_a J^a&=\\
     &e^{-\zeta}\Tilde{\nabla}_a \Tilde{J}^a    
     -4e^{-\zeta}(h^{ad} h^{bc}-h^{ac} h^{bd})\\
     [&-(\xi_{bd} \Tilde{K}_{ac})-(\xi_b\Tilde{\partial}_d+\xi_d\Tilde{\partial}_b) \Tilde{K}_{ac}\\
         &+\Tilde{\Gamma}^{p}_{ad}(\xi_b \Tilde{K}_{pc}+\xi_c \Tilde{K}_{pb}-\xi_p \Tilde{K}_{bc})\\
         &-\tau\{\xi_{bd}+(\xi_b\Tilde{\partial}_d+\xi_d\Tilde{\partial}_b)\}(\partial_\tau \Tilde{K}_{ac})\\
         &+\tau \Tilde{\Gamma}^p_{ad}(\xi_b \partial_\tau \Tilde{K}_{pc}+\xi_c \partial_\tau \Tilde{K}_{pb}-\xi_p\partial_\tau \Tilde{K}_{bc})\\
         &+\xi_b\xi_d \Tilde{K}_{ac}
         +3\tau \xi_b\xi_d\partial_\tau \Tilde{K}_{ac}+\xi_b\xi_d \tau^2\partial_\tau^2 \Tilde{K}_{ac}]\\
         &+\mathcal{O}(\epsilon^2).\label{eq2.20}
         \end{split}
 \end{equation}
    
From equations (\ref{eq2.14}) and (\ref{eq2.20}), we find that terms linear in $\Tilde{K}_{ab}$, i.e., $\mathcal{O}(\epsilon)$ terms cancel exactly leaving an overall factor of $e^{-\zeta}$ in the zeroth order term. Hence, we have
\begin{equation}
     \frac{1}{\sqrt{h}} \partial_v(\sqrt{h} J^v)+\nabla_a J^a=  e^{-\zeta}\bigg[ \frac{1}{\sqrt{h}} \partial_\tau\mathcal(\sqrt{h}J^\tau)+\Tilde{\nabla}_a \Tilde{J}^a\bigg]+\mathcal{O}(\epsilon^2).
\end{equation}
\section{Conclusion}\label{sec:conclude}
In this note, we have verified the general expectation that net entropy production in a dynamical gravity should not depend on how we choose coordinates along the horizon. First, in section \ref{sec:symmetry}, we have outlined a general proof of why the entropy production should transform in the way we physically expect (see equation \eqref{eq:predic} and the discussion around). Then in the next section, we verified the claim for the particular case of Gauss-Bonnet theory by explicit computation. This provides a  consistency check on the construction of the entropy current in Einstein-Gauss-Bonnet theory.

It might seem that apart from the consistency check mentioned above,  our computation is not of much use since we already have a general proof that this symmetry must work. However, as we have already mentioned in the introduction, our final goal is to have some construction of entropy current and entropy density that works without any perturbation.
In this context, it would be interesting to analyze this symmetry in a more systematic manner so that we could use it to constrain the structure of the entropy density and the entropy current in a theory-independent manner. Note that the existence of entropy density and the spatial entropy current has been predicted using the special case of the transformation considered here, namely boost symmetry generated by a constant $\zeta$\cite{Wall},\cite{our2105}. It is natural to expect more constraints in the whole structure if we use a larger symmetry where $\zeta$ is a function of all spatial coordinates.
Our note is a small step towards this goal which would give us more experience in dealing with the symmetries of null surfaces and corresponding transformation of the relevant physical quantities. 

One very natural extension of this work might be to perform similar calculations for other four-derivative theories where the cancellations can be slightly non-trivial due to the presence of off-the-horizon terms in the entropy current and entropy density.

Another interesting future direction to take can be to explore the existence of any possible relations between this reparametrization symmetry and the BMS or Carrollian symmetries. Recently in \cite{Donnay, Donnay2, Donnay3, BlauBMS}, the authors have shown the presence of extended BMS-like symmetries on the black hole horizon called Carrollian symmetries. Any possible connections of this symmetry with supertranslations or superrotations of the others can be useful in our understanding of the rich symmetric structure of the horizon.

\section{acknowledgments}
We would like to thank Parthajit Biswas, Anirban Dinda and Nilay Kundu for initial collaboration, several useful discussions and many important inputs. S.B. would like to thank Shiraz Minwalla, Sunil Mukhi and Yogesh K. Srivastava for many valuable discussions. We acknowledge funding from the Department of Atomic Energy, India. We would like to acknowledge our debt to the people of India for their steady and generous support to research in the basic sciences.

\appendix

\section{Notations, Conventions, and  Definitions}

In this appendix, we summarize our notation conventions and list the definitions of the various structures that we have used throughout our work.

\begin{itemize}
    \item Indices: Uppercase Latin alphabets $A,B,C...$ will refer to full $D$ space-time coordinates and Lowercase Latin alphabets $a,b,c...$ will refer to the $(D-2)$ dimensional spatial coordinates. 
    \item Choice of coordinates:
    \begin{equation*}
        \begin{split}
            X^A &= \{ r, v, x^a\},~~ Y^A = \{ \rho, \tau, y^a \} : \text{ The full space-time }\\
            &\text{coordinates in $D$ dimensions} \\
            r, \rho &= \text{ The radial coordinates} \\
            v, \tau &= \text{ The  Eddington-Finkelstein type time coordinates} \\
            x^a, y^a &=  \text{ The $(D-2)$ spatial coordinates}
        \end{split}
    \end{equation*}
    \item Choice of Space-time Metrics:
    \begin{equation*}
        \begin{split}
            ds^2 &= 2 ~dv~ dr -r^2 ~X(r,v,x^a)~dv^2 \\
            &+ 2~ r~\omega_a(r,v,x^b)~ dv~dx^a + h_{ab}(r,v,x^a)~dx^a~dx^b \\
            &= G_{AB}(r,v,x^a) ~dX^A~ dX^B \\
            &= 2~d\tau~d\rho - \rho^2~ \Tilde{X}(\rho,\tau,y^a)~d\tau^2 \\
            &+ 2~\rho~\Tilde{\omega}_a~(\rho,\tau,y^b)~d\tau~dy^a + \Tilde{h}_{ab}(\rho,\tau,y^a) ~dy^a~dy^b \\
            &= g_{AB}(\rho,\tau,y^a) ~dY^A~dY^B
        \end{split}
    \end{equation*}
    \item Structures like spatial derivatives, curvature tensors, and metric components in the $Y^A$ coordinate system will be represented with a  $\Tilde{~~}$ 
    on their corresponding counterparts in the $X^A$ coordinates. For example, $X, \omega_i, h_{ij}, \left(\partial_a=\frac{\partial}{\partial x^a}\right) \rightarrow \Tilde{X}, \Tilde{\omega}_i, \Tilde{h}_{ij}, \left( \Tilde{\partial}_a= \frac{\partial}{\partial y^a}\right) $
    \item Transformation of Coordinates and Derivatives on the Horizon:
    \begin{align*}
            r &= e^{-\zeta}\rho + \mathcal{O}(\rho^2) \\
            \rho &= e^\zeta r + \mathcal{O}(r^2)  \\
            v &= e^\zeta \tau + \mathcal{O}(\rho) \\
            \tau &= e^{-\zeta}v + \mathcal{O}(r)    \\
            x^a &= y^a + \mathcal{O}(\rho) \\
            y^a &= x^a + \mathcal{O}(r)  \\
            \partial_r &= e^\zeta \left(  \partial_\rho + \frac{1}{2} \tau^2 \xi^2 \partial_\tau + \tau \xi^a \Tilde{\partial}_a \right)  + \mathcal{O}(\rho) \\
            \partial_v &= e^{-\zeta}\partial_\tau + \mathcal{O}(\rho) \\
            \partial_a &= \Tilde{\partial}_a - \tau~ \xi_a \partial_\tau + \mathcal{O}(\rho)
    \end{align*}
    where we've denoted $\partial_a \zeta = \Tilde{\partial}_a \zeta$ by $\xi_a$.
    \item Definition of Curvature Tensors:
    \begin{align*}
        K_{ab} &= \frac{1}{2} \partial_v h_{ab} & K = h^{ab}K_{ab} &= \frac{1}{\sqrt{h}} \partial_v \sqrt{h} \\
        \Tilde{K}_{ab} &= \frac{1}{2} \partial_\tau \Tilde{h}_{ab} & \Tilde{K} = \Tilde{h}^{ab} \Tilde{K}_{ab} &= \frac{1}{\sqrt{\Tilde{h}}} \partial_\tau \sqrt{\Tilde{h}} 
    \end{align*}
    $$R_{ABCD}, R_{AB}, R =$$    Riemann tensor, Ricci tensor, Ricci scalar corresponding to full metric $G$  or  $g$
    $$ \mathcal{R}_{abcd}, \mathcal{R}_{ab}, \mathcal{R} =$$
    Riemann tensor, Ricci tensor, Ricci scalar corresponding to intrinsic metric $h$ or $\Tilde{h}$ 
\end{itemize}

\section{Detailed Expressions}
In this appendix, we show the explicit calculations for the relation between quantities such as Christoffel connection, Ricci scalar, and the divergence of entropy current  between $X^A$ and $Y^A$ coordinate systems.

\begin{itemize}
  
\item Expression for Christoffel connection in transformed coordinates 
\begin{equation}
    \begin{split}
        \Gamma_{a,bc} &=\frac{1}{2} \big(\partial_b  h_{ac}+\partial_c  h_{ab}-\partial_a h_{bc}\big)\\
       &=\Tilde{\Gamma}_{a,bc}-\frac{1}{2}\tau\partial_{\tau}\big(\xi_b h_{ac}+\xi_c h_{ab}-\xi_a h_{bc}\big)\\
       &= \Tilde{\Gamma}_{a,bc}-\tau\big(\xi_b \Tilde{K}_{ac}+ \xi_c \Tilde{K}_{ab}-\xi_a \Tilde{K}_{bc}\big)\label{Christoffel}
    \end{split}
\end{equation}
\item Expressions for Riemann tensor and Ricci scalar
 \begin{equation}
  \mathcal{R}_{abcd}=-[\partial_d \Gamma_{a,bc}-\partial_c \Gamma_{a,bd}+h^{pq}\Gamma_{p,ac} \Gamma_{q,bd}-\Gamma_{p,ad}\Gamma_{q,bc}h^{pq}]\label{Reimann}
\end{equation}

\item {Ricci Scalar in transformed coordinates}

     \begin{equation}
     \begin{split}
    \mathcal{R}&=h^{ac} h^{bd} R_{abcd}\\
    &=-h^{ac} h^{bd} \partial_d \Gamma_{a,bc}+h^{ad}h^{bc}\partial_d \Gamma_{a,bdc}\\
    &+h^{ac} h^{bd} \Gamma_{p,ad}\Gamma_{q,bc}h^{pq}-h^{ad}h^{bc}\Gamma_{p,ad}\Gamma_{q,bc}h^{pq}\\
    &=(h^{ad}h^{bc}-h^{ac}h^{bd})\big(\partial_d \Gamma_{a,bc}-h^{pq}\Gamma_{p,ad}\Gamma_{q,bc}\big)\label{Ricci}
\end{split}
\end{equation}

\begin{equation}
   \begin{split}
        \partial_d  \Gamma_{a,bc} =&\Tilde{\partial}_d \Gamma_{a,bc} -\tau\xi_d\partial_\tau\Gamma_{a,bc} \\
        =&\Tilde{\partial}_d \big[\Tilde{\Gamma} _{a,bc}-\tau(\xi_b \Tilde{K}_{ac}+\xi_c \Tilde{K}_{ab}-\xi_a \Tilde{K}_{bc})\big]\\
        &-\tau\xi_d\partial_\tau\big[\Tilde{\Gamma} _{a,bc}-\tau(\xi_b \Tilde{K}_{ac}+\xi_c \Tilde{K}_{ab}-\xi_a \Tilde{K}_{bc})\big]\\
        =&\big[\Tilde{\partial}_d \Tilde{\Gamma} _{a,bc}- \tau(\xi_{bd} \Tilde{K}_{ac}+\bcancel{\xi_{cd} \Tilde{K}_{ab}}-\xi_{ad} \Tilde{K}_{bc})\\
        &-\tau(\xi_b\Tilde{\partial}_d \Tilde{K}_{ac}+\bcancel{\xi_c \Tilde{\partial}_d \Tilde{K}_{ab}}-\xi_a \Tilde{\partial}_d \Tilde{K}_{bc})\big]\\
        &+\big[-\tau  (\xi_d\Tilde{\partial}_b \Tilde{K}_{ac}+\bcancel{\xi_d\Tilde{\partial}_c \Tilde{K}_{ab}}-\xi_d\Tilde{\partial}_a \Tilde{K}_{bc})\\
        &+\tau (\xi_d\xi_b \Tilde{K}_{ac}+\bcancel{\xi_d\xi_c \Tilde{K}_{ab}}-\xi_a\xi_d \Tilde{K}_{bc})\\
        &+\tau^2(\xi_d\xi_b\partial_\tau \Tilde{K}_{ac}+\bcancel{\xi_d\xi_c \partial_\tau \Tilde{K}_{ab}}-\xi_d\xi_a\partial_\tau \Tilde{K}_{bc})\big].\label{Ricci_trans1}
      \end{split}
\end{equation}

The terms canceled in (\ref{Ricci_trans1})  due to the fact that terms symmetric in $(c,d)$ will not contribute to the Ricci scalar as it has a prefactor of $(h^{ad}h^{bc}-h^{ac}h^{bd})$ which is antisymmetric in $(c,d)$. Hence, 
\begin{equation}
\begin{split}
     \partial_d  \Gamma_{a,bc} =\Tilde{\partial}_d \Tilde{\Gamma} _{a,bc}- \tau\big[&(\xi_{bd}+\xi_b\Tilde{\partial}_d+\xi_d\Tilde{\partial}_b-\xi_d\xi_b)\Tilde{K}_{ac}\\
     &-(\xi_{ad}+\xi_a\Tilde{\partial}_d+\xi_d\Tilde{\partial}_a-\xi_a\xi_d) \Tilde{K}_{bc}]\\
     +\tau^2&[\xi_d\xi_b\partial_\tau \Tilde{K}_{ac}-\xi_d\xi_a \partial_\tau \Tilde{K}_{bc}]\label{Christoffel_derv}.
\end{split}
\end{equation}
From eqn. (\ref{Ricci}) and (\ref{Christoffel}),
\begin{equation}
\begin{split}
     h^{pq}\Gamma_{p,ad} \Gamma_{q,bc} &=h^{pq}\Tilde{\Gamma}_{p,ad} \Tilde{\Gamma}_{q,bc} -\tau h^{pq}\Tilde{\Gamma}_{p,ad} (\xi_b \Tilde{K}_{qc}+\xi_c \Tilde{K}_{qb}-\xi_q \Tilde{K}_{bc})\\
    &-\tau h^{pq}\Tilde{\Gamma}_{q,bc} (\xi_a \Tilde{K}_{pd}+\xi_d \Tilde{K}_{pa}-\xi_p \Tilde{K}_{ad})+\mathcal{O}(\epsilon^2)\label{Christoffel_prodct}.
\end{split}
   \end{equation}
Thus, from (\ref{Ricci}), (\ref{Christoffel_derv}), and (\ref{Christoffel_prodct})
\begin{equation}
\begin{split}
     \mathcal{R} &=\Tilde{\mathcal{R}} +(h^{ad}h^{bc}-h^{ac}h^{bd})\\
     &[-2\tau\{\xi_{bd}+(\xi_b\Tilde{\partial}_d+\xi_d\Tilde{\partial}_b)-\xi_d\xi_b\}\Tilde{K}_{ac}\\
     &+2\tau^2(\xi_d\xi_b\partial_\tau \Tilde{K}_{ac})+ 2\tau h^{pq}\ \Tilde{\Gamma}_{p,ad} (\xi_b \Tilde{K}_{qc}+\xi_c \Tilde{K}_{qb}-\xi_q \Tilde{K}_{bc})]\\
     &+\mathcal{O}(\epsilon^2)\label{RicciTrans}.
\end{split}
   \end{equation}

   \item{The divergence of entropy current in transformed coordinates}
  
   The expression for entropy current for Gauss-Bonnet theory is given as
   \begin{equation}
       J^{a}=-4(\nabla_b K^{ba}-\nabla^a K_{cd} h^{cd})\label{ED}.
   \end{equation}
   This implies, that the divergence of Entropy current is 
   \begin{equation}
   \begin{split}
                         \nabla_a J^{a}&=-4(h^{ad} h^{bc}-h^{cd} h^{ab})\nabla_a\nabla_b K_{cd}\\
            &=-4(h^{ad} h^{bc}-h^{ac} h^{bd})\nabla_b \nabla_d K_{ac}.
      \label{ED Div}
   \end{split}
   \end{equation}
     Let us define a three index object $M_{d,ac}$ such that 
 \begin{widetext}    
   \begin{equation}
   \begin{split}
          M_{d,ac} \equiv&\nabla_d  K_{ac} 
       =\nabla_d \big(e^{-\zeta} \Tilde{K}_{ac}\big)
       =\partial_d \big(e^{-\zeta}\Tilde{K}_{ac}\big)-\Gamma_{da}^p \big(e^{-\zeta} \Tilde{K}_{pc}\big)-\Gamma_{dc}^p \big(e^{-\zeta} \Tilde{K}_{ap}\big)\\
       =&\{\Tilde{\partial}_d -\xi_d\tau\partial_\tau\}\big(e^{-\zeta} \Tilde{K}_{ac}\big) -\Tilde{\Gamma}_{da}^{p}\big(e^{-\zeta} \Tilde{K}_{pc}\big)- \Tilde{\Gamma}_{dc}^{p}\big(e^{-\zeta} \Tilde{K}_{ap}\big) +\mathcal{O}(\epsilon^2)\\
       =& e^{-\zeta}\big[\Tilde{\partial}_d  \Tilde{K}_{ac}-\xi_d \Tilde{K}_{ac}- \xi_d\tau\partial_\tau \Tilde{K}_{ac} -\Tilde{\Gamma}_{da}^p \Tilde{K}_{pc}-\Tilde{\Gamma}_{dc}^p \Tilde{K}_{ap}\big]+\mathcal{O}(\epsilon^2)\\
       =&e^{-\zeta}\left(\Tilde{\nabla}_d  \Tilde{K}_{ac}-\xi_d \Tilde{K}_{ac}-\xi_d \tau\partial_\tau \Tilde{K}_{ac}\right)+\mathcal{O}(\epsilon^2)\\
       =& e^{-\zeta} \left(\Tilde{M} _{d,ac}-(\delta \Tilde{M})_{d,ac}\right)+\mathcal{O}(\epsilon^2)\label{M_trans},\\
   \end{split}
   \end{equation}
   
   where,
   
   \begin{equation}
       (\delta \Tilde{M})_{d,ac}= \xi_d \left(\Tilde{K}_{ac}+\tau\partial_\tau \Tilde{K}_{ac}\right)\label{deltaM}
   \end{equation}
   
  Also, we define

   \begin{equation}
   \begin{split}
              W_{abcd} \equiv&\nabla_b  M _{d,ac}\\
       =& \partial_b M_{d,ac} -\Gamma_{bd}^p  M_{p,ac} -\Gamma_{ba}^p  M_{d,pc} -\Gamma_{bc}^p  M_{d,ap} \\
       =& \Tilde{\partial}_b\left\{e^{-\zeta} \left(\Tilde{M}_{d,ac}-(\delta \Tilde{M})_{d,ac}\right)\right\}
       -\xi_b e^{-\zeta}\tau\partial_\tau{\left(\Tilde{M}_{d,ac}-(\delta \Tilde{M})_{d,ac}\right)}\\
       &-e^{-\zeta}\bigg[\Tilde{\Gamma}_{bd}^p\left(\Tilde{M}_{p,ac}-(\delta \Tilde{M})_{p,ac}\right)
       +\Tilde{\Gamma}_{ba}^p\left(\Tilde{M}_{d,pc}-(\delta \Tilde{M})_{d,pc}\right) +\Tilde{\Gamma}_{bc}^p\left(\Tilde{M}_{d,ap}-(\delta \Tilde{M})_{d,ap}\right)\bigg]+\mathcal{O}(\epsilon^2)\\
       =& e^{-\zeta}\left[\Tilde{\nabla}_b \Tilde{M}_{d,ac}-\xi_b (1+\tau\partial_\tau)\left(\Tilde{M}_{d,ac}-\delta \Tilde{M}_{d,ac}\right)-\Tilde{\nabla}_b \delta \Tilde{M}_{d,ac}\right]\\
       =& e^{-\zeta} \left[\Tilde{\nabla}_b\Tilde{\nabla}_d \Tilde{K}_{ac}-\xi_b \Tilde{M}_{d,ac}-\Tilde{\nabla}_b\delta \Tilde{M}_{d,ac}-\xi_b\tau\partial_{\tau}\Tilde{M}_{d,ac}+\xi_b (1+\tau\partial_\tau)\delta \Tilde{M}_{d,ac}\right]\\
       =&e^{-\zeta}\bigg[\Tilde{\nabla}_b\Tilde{\nabla}_d \Tilde{K}_{ac}\underbrace{-\xi_b \Tilde{\nabla}_d \Tilde{K}_{ac}-\Tilde{\nabla}_b (\xi_d \Tilde{K}_{ac})}_{\text{term 1}}-\underbrace{\Tilde{\nabla}_b (\xi_d\tau\partial_\tau \Tilde{K}_{ac})-\xi_b\tau\partial_\tau\Tilde{\nabla}_d \Tilde{K}_{ac}}_{\text{term 2}} +\underbrace{\xi_b (1+\tau\partial_\tau)(\xi_d \Tilde{K}_{ac}+\xi_d\tau\partial_\tau \Tilde{K}_{ac})}_{\text{term 3}}\bigg]+\mathcal{O}(\epsilon^2)\label{Wtrans}.
    \end{split}
   \end{equation}
\end{widetext}   
   
   Now, 
   \begin{equation}
   \begin{split}
       \text{ term 1}=&-\xi_b \Tilde{\nabla}_d \Tilde{K}_{ac}-\Tilde{\nabla}_b (\xi_d \Tilde{K}_{ac})\\
       =&-\xi_b \Tilde{\partial}_d \Tilde{K}_{ac}+\xi_b \Tilde{\Gamma}_{da}^p \Tilde{K}_{pc}+\xi_b \Tilde{\Gamma}_{dc}^p \Tilde{K}_{ap}\\
       &-\xi_d \Tilde{\partial}_b \Tilde{K}_{ac}+\xi_d \Tilde{\Gamma}_{ba}^p \Tilde{K}_{pc}+\xi_d \Tilde{\Gamma}_{bc}^p \Tilde{K}_{ap}\\
       &-\xi_{bd} \Tilde{K}_{ac}+\Tilde{\Gamma}_{bd}^p \xi_p \Tilde{K}_{ac}.\label{term1}
        \end{split}
         \end{equation}
         From (\ref{ED Div}), we see that for calculation of the divergence of entropy current, the terms in (\ref{Wtrans}) have to be contracted with $(h^{ad} h^{bc}-h^{ac} h^{bd})$, which is antisymmetric in $(c,d)$ or $(a,b)$. Now, in (\ref{term1}), the terms $\xi_b \Tilde{\Gamma}_{dc}^p \Tilde{K}_{ap}$  and $\xi_d \Tilde{\Gamma}_{ba}^p \Tilde{K}_{pc}$ are symmetric in $(c,d)$ and $(a,b)$ respectively. Hence these can be dropped. In addition, we can perform some relabelling of indices and rewrite term 1 as
         
         \begin{equation}
             \begin{split}
                 \text{term 1}=&-\xi_b \Tilde{\partial}_d \Tilde{K}_{ac}+\xi_b \Tilde{\Gamma}_{da}^p \Tilde{K}_{pc}\\
                 &-\xi_d \Tilde{\partial}_b \Tilde{K}_{ac}+\xi_c \Tilde{\Gamma}_{ad}^p \Tilde{K}_{bp}\\
                 &-\xi_{bd} \Tilde{K}_{ac}-\Tilde{\Gamma}_{ad}^p \xi_p \Tilde{K}_{bc}.\\
                 =&-(\xi_b\Tilde{\partial}_d+\xi_d\Tilde{\partial}_b) \Tilde{K}_{ac}-\xi_{bd} \Tilde{K}_{ac}\\
                 &+\Tilde{\Gamma}_{ad}^p (\xi_c \Tilde{K}_{pb}+\xi_b \Tilde{K}_{pc}-\xi_p \Tilde{K}_{bc})\label{term1a}.
             \end{split}
         \end{equation}
         
       In a similar fashion, we can express term 2 as 
       
       \begin{equation}
   \begin{split} 
         \text{term 2} =& -\xi_b\tau\partial_\tau\Tilde{\nabla}_d \Tilde{K}_{ac}-\Tilde{\nabla}_b (\xi_d\tau\partial_\tau \Tilde{K}_{ac})\\
         =&-\tau\left[\xi_b\Tilde{\nabla}_d({\partial}_\tau \Tilde{K}_{ac})+\Tilde{\nabla}_b (\xi_d{\partial}_\tau \Tilde{K}_{ac})\right]+\mathcal{O}(\epsilon^2)\\
         =& -\tau \Big[(\xi_b\Tilde{\partial}_d+\xi_d\Tilde{\partial}_b) \partial_{\tau}\Tilde{K}_{ac}+\xi_{bd} \partial_{\tau}\Tilde{K}_{ac}\\
                 &-\Tilde{\Gamma}_{ad}^p (\xi_c \partial_{\tau}\Tilde{K}_{pb}+\xi_b \partial_{\tau}\Tilde{K}_{pc}-\xi_p \partial_{\tau}\Tilde{K}_{bc})\Big]+\mathcal{O}(\epsilon^2)\label{term2}.
   \end{split}
         \end{equation}
         Now, evaluating term 3
         \begin{equation}
             \begin{split}
                 \text{term 3}=&\xi_b (1+\tau\partial_\tau)(\xi_d \Tilde{K}_{ac}+\xi_d\tau\partial_\tau \Tilde{K}_{ac})\\
                 =& \xi_b\xi_d(\Tilde K_{ac}+3\tau\partial_\tau \Tilde K_{ac}+\tau^2\partial_\tau^2 \Tilde K_{ac})\label{term3}.
             \end{split}
         \end{equation}
         Combining results from (\ref{Wtrans}), (\ref{term1a}), (\ref{term2}) and (\ref{term3})

         \begin{equation}
             \begin{split}
                 W_{abcd}=e^{-\zeta}\bigg[&\Tilde{\nabla}_b\Tilde{\nabla}_d \Tilde{K}_{ac}-\left(\xi_{bd} \Tilde{K}_{ac}\right)-\left(\xi_b\Tilde{\partial}_d+\xi_d\Tilde{\partial}_b\right) \Tilde{K}_{ac}\\
         &+\Tilde{\Gamma}^{p}_{ad}\left(\xi_b \Tilde{K}_{pc}+\xi_c \Tilde{K}_{pb}-\xi_p \Tilde{K}_{bc}\right)\\
         &-\tau\left\{\xi_{bd}+\left(\xi_b\Tilde{\partial}_d+\xi_d\Tilde{\partial}_b\right)\right\}(\partial_\tau \Tilde{K}_{ac})\\
         &+\tau \Tilde{\Gamma}^p_{ad}\left(\xi_b \partial_\tau \Tilde{K}_{pc}+\xi_c \partial_\tau \Tilde{K}_{pb}-\xi_p\partial_\tau \Tilde{K}_{bc}\right)\\
         &+\xi_b\xi_d \Tilde{K}_{ac}
         +3\tau \xi_b\xi_d\partial_\tau \Tilde{K}_{ac}+\xi_b\xi_d \tau^2\partial_\tau^2 \Tilde{K}_{ac}\bigg]\\
         &+\mathcal{O}(\epsilon^2).
             \end{split}
         \end{equation}

         Hence, the divergence of entropy current becomes

         \begin{equation}
             \begin{split}
                 \nabla_a J^a&=e^{-\zeta}\Tilde{\nabla}_a \Tilde{J}^a    
     -4e^{-\zeta}(h^{ad} h^{bc}-h^{ac} h^{bd})\\
        [&-(\xi_{bd} \Tilde{K}_{ac})-(\xi_b\Tilde{\partial}_d+\xi_d\Tilde{\partial}_b) \Tilde{K}_{ac}\\
         &+\Tilde{\Gamma}^{p}_{ad}(\xi_b \Tilde{K}_{pc}+\xi_c \Tilde{K}_{pb}-\xi_p \Tilde{K}_{bc})\\
         &-\tau\{\xi_{bd}+(\xi_b\Tilde{\partial}_d+\xi_d\Tilde{\partial}_b)\}(\partial_\tau \Tilde{K}_{ac})\\
         &+\tau \Tilde{\Gamma}^p_{ad}(\xi_b \partial_\tau \Tilde{K}_{pc}+\xi_c \partial_\tau \Tilde{K}_{pb}-\xi_p\partial_\tau \Tilde{K}_{bc})\\
         &+\xi_b\xi_d \Tilde{K}_{ac}
         +3\tau \xi_b\xi_d\partial_\tau \Tilde{K}_{ac}+\xi_b\xi_d \tau^2\partial_\tau^2 \Tilde{K}_{ac}]\\
         &+\mathcal{O}(\epsilon^2).
             \end{split}
         \end{equation}
           
\end{itemize}

\section{Action of Derivatives on some Specific Structures}

    In this appendix we'll see how the derivatives of certain boost weight $1$ structures transform under the coordinate transformations. We'll see how these terms can be condensed into some particular forms that can help us manipulate them in simpler ways.

    Any boost weight $1$ term can be written in the form of $\partial_v$(some boost weight 0 structure, say $Q_{a_1 a_2...a_n}$). Transforming the $\partial_v$ operator under the coordinate transformations as in \eqref{eq2.6}, we can write it as $e^{-\zeta} (\partial_\tau Q_{a_1 a_2 ... a_n})$. Also since $\tau$ is analogous to the $v$ coordinate itself, $(\partial_\tau Q_{a_1 a_2 ... a_n})$ itself is a boost weight $1$ structure in the $\{\rho,\tau,y^a\}$ coordinate system. Now if we act with a $\nabla_{x^i}$ on this structure, we get

    \begin{equation}
    \begin{split}
        \nabla_{i} (\partial_v  Q_{a_1 a_2...a_n}) &= \nabla_{i}(e^{-\zeta} (\partial_\tau Q_{a_1 a_2...a_n})) \\
        &= \partial_{i} (e^{-\zeta} (\partial_\tau Q_{a_1 a_2...a_n})) - e^{-\zeta} \Gamma^b_{i a_1} \partial_\tau Q_{b a_2...a_n} \\
        &- e^{-\zeta} \Gamma^b_{i a_2} \partial_\tau Q_{a_1 b ...a_n} ...\\ 
        &- e^{-\zeta} \Gamma^b_{i a_n} \partial_\tau Q_{a_1 a_2...b} 
    \end{split}
    \end{equation}

    \begin{equation}
    \begin{split}
        &\partial_{i} (e^{-\zeta} (\partial_\tau Q_{a_1 a_2...a_n})) \\
        &= (\Tilde{\partial}_i - \xi_i \tau \partial_\tau)  (e^{-\zeta} (\partial_\tau Q_{a_1 a_2...a_n})) \\
        &= -\xi_i  (e^{-\zeta} (\partial_\tau Q_{a_1 a_2...a_n})) - \xi_i \tau (e^{-\zeta} \partial_\tau (\partial_\tau Q_{a_1 a_2...a_n})) \\
        &- e^{-\zeta} \Tilde{\partial}_i  (\partial_\tau Q_{a_1 a_2...a_n}) \\
        &=  e^{-\zeta} \left[\Tilde{\partial}_i  (\partial_\tau Q_{a_1 a_2...a_n})-  \xi_i (1+\tau \partial_\tau)  (\partial_\tau Q_{a_1 a_2...a_n}) \right] \\
        &\Gamma^b_{i a_m}  (\partial_\tau Q_{a_1 a_2..b..a_n}) = \left[\Tilde{\Gamma}^b_{i a_m} - \tau (\xi \Tilde{K} ...) \right]  (\partial_\tau Q_{a_1 a_2..b..a_n}) \\
        &= \Tilde{\Gamma}^b_{i a_m}   (\partial_\tau Q_{a_1 a_2..b..a_n}) + \mathcal{O}(\epsilon^2) \\
        \Rightarrow &\nabla_i ( \partial_v Q_{a_1 a_2...a_n}) = e^{-\zeta} \left[ \Tilde{\nabla}_i - \xi_i (1+\tau \partial_\tau) \right] \partial_\tau Q_{a_1 a_2...a_n} + \mathcal{O}(\epsilon^2)
    \end{split}
\end{equation}

This form becomes especially useful while calculating $J^i$ and $\nabla_i J^i$.

One more structure that can appear in the calculations of the $\partial_v J^v$ is of the form $\partial_v (\tau Q)$ from the extra terms that are generated due to the coordinate transformation. This derivative can be arranged in the following form which makes it easier to manipulate.
\begin{equation}
    \begin{split}
        \partial_v \left[ \tau  Q \right] &= e^{-\zeta}\partial_\tau \left[ \tau  Q \right] = e^{-\zeta} (1+\tau \partial_\tau) Q
    \end{split}
\end{equation}


%\bibliography{GBEC}% Produces the bibliography via BibTeX.

\end{document}